\documentclass[fleqn,twoside]{article}
\usepackage{espcrc2}
\usepackage{epsfig}

\usepackage{graphicx}
\usepackage[figuresright]{rotating}

\newcommand{\AmS}{{\protect\the\textfont2
  A\kern-.1667em\lower.5ex\hbox{M}\kern-.125emS}}

\title{Vector meson production and inclusive $K^0_sK^0_s$ final state at HERA}

\author{M. Barbi
\address{McGill University, 
        Physics Department\\
        Montreal, Quebec\\ 
	Canada}
	\\ (on behalf of the H1 and ZEUS Collaborations)}

\begin{document}

\begin{abstract}
Measurements of vector mesons at HERA allow a detailed study of diffractive
and non-diffractive production mechanisms, fragmentation and decay branching
ratios. Results are  presented on hadronic resonance measurements.
In addition, the first observation of two meson states at masses around 1500 MeV 
and 1700 MeV are reported using inclusive $K^0_sK^0_s$ production
in deep inelastic $ep$ scattering in ZEUS at HERA.
\vspace{1pc}
\end{abstract}

% typeset front matter (including abstract)
\maketitle

\section{Vector Meson Production at HERA}

Production of light and heavy vector mesons 
(V) in exclusive and in proton dissociation processes 
has been studied in
$ep$ reactions at HERA in a wide
range of the $\gamma p$ centre-of-mass energy $W$ 
and photon virtuality $Q^2$. For very low $Q^2$ and $W> 10$ GeV, 
these reactions display features characteristic
of a soft diffractive process. The cross sections rises weakly with 
the energy $W$, \mbox{$\sigma_{V}(W) \propto W^\delta$} ($\delta\simeq 0.22$),
and has a steep exponential $t$ dependence, 
%\mbox{$d\sigma_{V}/{dt} \propto e^{-b(W)|t|}$}, 
where $t$ is the squared 
four-momentum transfer at the proton vertex. Such
processes are well described within the framework of 
Regge phenomenology~\cite{regge1,regge2,regge3} and the Vector-Meson Dominance model 
(VDM)~\cite{vdm1,vdm2},  where 
%exclusive vector-meson (VM)  
%production is assumed to proceed  
the photon is assumed to fluctuate into a vector
meson before scattering from the proton via Pomeron exchange.
However, this approach fails at high values of $Q^2$ or
high vector meson masses.
In the hard regime, models based on perturbative QCD (pQCD) 
\cite{zfp:c57:89,pr:d50:3134}
can be used to describe the vector meson production. The photon fluctuates into 
a $q\bar q$ state and the quark dipole interacts with the proton in the 
lowest order via two gluons exchange 
\cite{zfp:c68:137,pr:d54:5523,pl:b478:101,hep-ph-0107068}. 
The exchange of the 
gluon ladder has also been calculated in the 
leading logarithm approximation (LLA) 
\cite{zfp:c68:137,hep-ph-0107068,pr:d53:3564,pl:b375:301}. 
The cross section is related to the rise of the gluon 
density in the proton as $x$ decreases,
%($x\sim 1/W^2$), 
%$\sigma_{VM}(W) \propto [xg(x,Q^2_{eff})]^2$, 
where $x$ is 
the Bjorken scale variable, 
%(**ref olhar DESY-02-072**) and $Q^2_{eff}$ is a 
%hard scale, 
and has a strong W dependence, 
$\sigma_{V}(W) \propto W^\delta$ ($\delta\simeq 0.8$.)

In this contribution, perturbative QCD models are compared 
with results from vector meson production in photoproduction and deep inelastic 
scattering at HERA.

\subsection{Exclusive Vector Meson Production}

\begin{figure}[htb]
%\framebox[55mm]{\includegraphics{wxsecbw.eps}}{\rule[-21mm]{0mm}{43mm}}
\epsfig{file=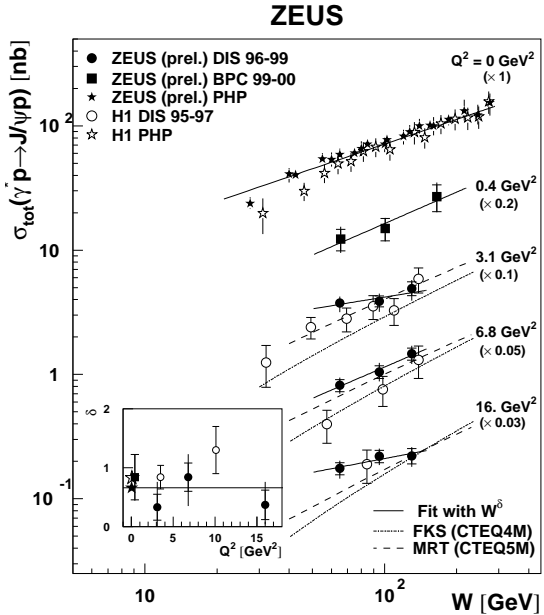,width=0.5\textwidth}
\caption{W-dependence of $J/\psi$ electroproduction at different values
of $Q^2$, and comparison with predictions from FKS and MRT models. The inset
displays the fitted value of $\delta$ as a function of $Q^2$.}
\label{fig:1}
\end{figure}

Figure~\ref{fig:1} shows $\sigma_{tot}^{\gamma^*p\rightarrow J/\psi p}$
from ZEUS \cite{zeus:jpsidis1,zeus:jpsidis2} and H1 \cite{h1:jpsidis} as a 
function of W for fixed $Q^2$, and also in photoproduction. 
The data is fitted with $W^\delta$. The ZEUS measurements are shown in 
table~\ref{tab1}. The results are compatible with results in photoproduction 
from H1 \cite{h1jpsiq2} $\delta =0.83\pm 0.07(stat\oplus syst)$
and ZEUS \cite{zeusjpsiq2} $\delta =0.69\pm 0.02(stat)\pm 0.03(syst)$. 
The cross sections are also compared to pQCD predictions 
from Frankfurt, Koepf and Strikman (FKS) 
\cite{fks} using CTEQ4M PDF \cite{cteq4m}, and from Martin, Ryskin and
Teubner (MRT) \cite{mrt} using CTEQ5M PDF \cite{cteq5m}. 
The results are sensitive to the PDF used and have 
large theoretical uncertainties, but they qualitatively 
describe the strong W-dependence of the cross section. 
They also indicate that at high vector meson masses $M_V$, a hard
regime already exist at very low values of $Q^2$, and 
that $M_V^2$ may set a hard scale.

\begin{table*}[htb]
\caption{Results from the fit of $W^\delta$
to $\sigma_{tot}^{\gamma^*p\rightarrow J/\psi p}$ measured 
by ZEUS at different values of $Q^2$ .}
\label{tab1}
\newcommand{\m}{\hphantom{$-$}}
\newcommand{\cc}[1]{\multicolumn{1}{c}{#1}}
\renewcommand{\tabcolsep}{.2pc} % enlarge column spacing
\renewcommand{\arraystretch}{1.2} % enlarge line spacing
\begin{tabular}{@{}lllll}
\hline
$Q^2$ (GeV$^2$) & \cc{$0.4$} & \cc{$3.2$} & \cc{$6.8$} & \cc{$16$} \\
\hline
$\delta$ & \m$0.84\pm 0.38^{+0.10}_{-0.04}$ & 
\m$0.33\pm 0.22^{+0.37}_{-0.46}$ & 
\m$0.84\pm 0.24^{+0.31}_{-0.43}$ & 
\m$0.37\pm 0.25^{+0.30}_{-0.22}$ \\
\hline
\end{tabular}\\[2pt]
\end{table*}

%\begin{figure}[htb]
%\framebox[55mm]{\includegraphics{wxsecbw.eps}}{\rule[-21mm]{0mm}{43mm}}
%\epsfig{file=jpsi_4xsec_mrtfks-c.eps,width=0.45\textwidth}
%\caption{Remember to keep details clear and large enough.}
%\label{fig:1}
%\end{figure}

\begin{figure}[htb]
%\framebox[55mm]{\includegraphics{wxsecbw.eps}}{\rule[-21mm]{0mm}{43mm}}
\epsfig{file=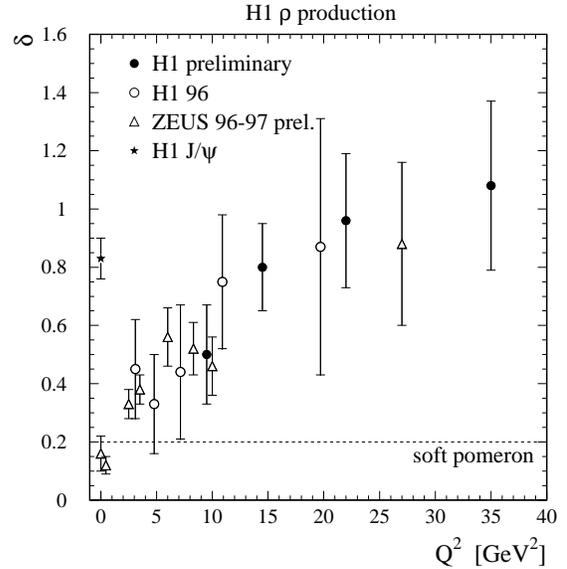,width=0.5\textwidth}
\caption{Results from the fit using $\sigma_{V}(W) \propto W^\delta$ for
the light and heavy vector mesons $\rho^0$ and $J/\psi$, respectively.
At low values of $Q^2$, soft processes are dominant for $\rho^0$
production, while hard processes dominate for $J/\psi$ production
already at $Q^2=0$.}
\label{fig:2}
\end{figure}

The $Q^2$-dependence of the $\delta$ parameter 
%results of the fit to the cross section  
%using the functional $W^\delta$ 
is shown in Fig. \ref{fig:2} for the light vector meson $\rho^0$ from 
ZEUS \cite{rho:zeus:eps01:594} and H1 \cite{rho:h1:ichep02:989}, 
and for $J/\psi$ in photoproduction from H1 \cite{h1jpsiq2}. 
At very low values of $Q^2$, Regge phenomenology and 
VDM models describe the $\rho^0$ meson production. However,
the energy W-dependence becomes steeper at high $Q^2$, 
%and hard processes start
%to dominate at \mbox{$Q^2>10$}~GeV$^2$, 
with a smooth transition from soft to hard
regime. On the other hand, the $J/\psi$ cross section has a strong 
energy dependence 
already at $Q^2=0$. These results indicate that $Q^2$ may also set a 
hard scale. 

\subsection{Vector Meson Production with Proton Dissociation}

\begin{figure}[htb]
%\framebox[55mm]{\includegraphics{wxsecbw.eps}}{\rule[-21mm]{0mm}{43mm}}
\epsfig{file=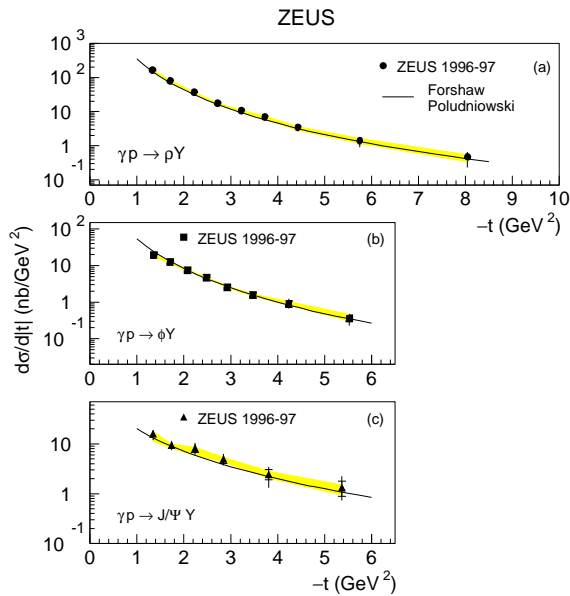,width=0.52\textwidth}
\caption{The differential cross sections
  ${\rm d}\sigma_{\gamma p \to V Y}/{\rm d}|t|$  
  for: (a) $\rho^0$, (b) $\phi$, and (c) $J/\psi$ photoproduction.
  The solid lines show the predictions from Forshaw and 
  Poludniowski using LLA BFKL.
  The inner bars indicate
  the statistical uncertainty and the outer bars represent the statistical and
  systematic uncertainties added in quadrature. The
  shaded bands represents the uncertainties due 
  to the  modelling of the hadronic-system $Y.$}
\label{fig:3}
\end{figure}

It was predicted \cite{prl:63:1914,pl:b284:123} that hard processes
may occur at high $-t$. 
Typical elastic vector meson production has \mbox{$-t<1$ GeV$^2$}, 
and cannot be used to verify these predictions. On the other hand, 
proton-dissociation processes have higher values of $-t$.
Figure \ref{fig:3} shows the differential cross sections of $\rho$, $\phi$
and $J/\psi$ vector mesons in the photoproduction proton-dissociative
reaction $\gamma p\rightarrow VY$ in the range \mbox{$80<W<120$} GeV and
\mbox{$x>0.01$} from
ZEUS \cite{eur:phys:101140}, 
where $Y$ is the dissociated hadronic 
system. The data is compared to predictions from Forshaw and 
Poludniowski~\cite{hep-ph-0107068} using LLA BFKL \cite{bfkl1,bfkl2}. 
The results are in good agreement, indicating that pQCD can describe
vector meson production at high $-t$ regime.
 
\subsection{Summary}

The results presented in this contribution indicates 
a smooth transition from soft to hard regime. They also show 
that pQCD describes the vector production in the hard regime, 
and that $Q^2$, $M_V^2$ and $-t$ can set a hard scale.

\section{Inclusive $K^0_sK^0_s$ Final State in Deep Inelastic Scattering at HERA}

The $K_s^0K_s^0$ system is expected to couple to scalar
and tensor glueballs. This has motivated intense 
experimental and theoretical study during the past few years
\cite{rmp:v71:1411,hep-ex/0101031}.
Lattice QCD calculations \cite{glueball1,glueball2}
predict the existence of a scalar 
glueball with a mass of $1730\pm 100$ MeV and 
a tensor glueball at $2400\pm 120$ MeV.
The scalar glueball can mix with $q\overline{q}$ states with $I=0$ from 
the scalar meson nonet, leading to three $J^{PC}=0^{++}$ states whereas
only two can fit into the nonet.
Experimentally, four states with $J^{PC}=0^{++}$ and $I=0$ have been 
established \cite{pdg}: $f_0(980)$, $f_0(1370)$, $f_0(1500)$ and $f_0(1710)$.

The state most frequently considered to be a glueball candidate 
is $f_0(1710)$ \cite{pdg} , but its gluon content has not 
yet been established. 
This state was first observed in radiative $J/\psi$ 
decays \cite{BES96} and
its angular momentum $J=0$ was established by the WA102 experiment 
using a partial-wave analysis in the $K^+K^-$ and $K_s^0K_s^0$ 
final states\cite{WAf1710} .
Observation of $f_0(1710)$ in $\gamma\gamma$ 
collisions may indicate a large quark content.
A recent publication from L3 \cite{L3} 
reports the observation of two states in $\gamma\gamma$ collisions
above $1500$ MeV, the well-established $f_2^\prime(1525)$ \cite{pdg}
and a broad resonance at 1760 MeV. 
It is not clear if the latter state is the $f_0(1710)$.
% or a radially excited state. 

The $ep$ collisions at HERA provide an opportunity to study resonance
production in a new environment. Production of $K_s^0$ particles 
has been studied previously at HERA \cite{epj:c2:77,zfp:c68:29,np:b480:3}.
In this contribution, the first observation of resonances in the 
$K_s^0K_s^0$ final state in inclusive deep inelastic $ep$ scattering 
(DIS) is reported \cite{paganis}.

\subsection{Results}

The data used for this study correspond to a total integrated luminosity of 
120 pb$^{-1}$ collected in ZEUS during the 1996-2000 running period. 

Oppositely charged track 
pairs reconstructed by the ZEUS central tracking detector (CTD) and 
assigned to a secondary vertex were selected and combined 
to form $K_s^0$ candidates. 
Both tracks were assigned the mass of a charged pion and the 
invariant-mass $M(\pi^+\pi^-)$ was calculated. 
Only events with at least one pair of $K_s^0$ candidates were selected.
The invariant mass of the $K_s^0$ pair candidate $M(K_s^0,K_s^0)$
was reconstructed in the range \mbox{$0.995~<~M(K_s^0K_s^0)~<2.795$ GeV}.

\begin{figure}[htb]
%\framebox[55mm]{\includegraphics{wxsecbw.eps}}{\rule[-21mm]{0mm}{43mm}}
\epsfig{file=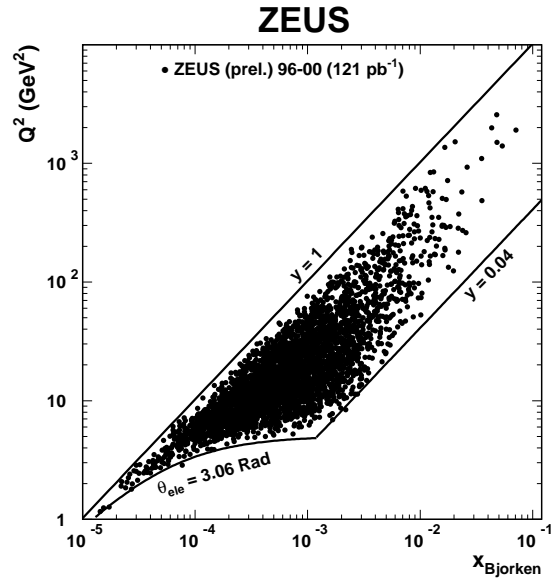,width=0.5\textwidth}
\caption{The distribution in $x$ and $Q^2$ of events passing all selection cuts.
The $y=0.04$ and $\theta_{ele}=3.1$ rad lines delineate approximately the 
kinematic region selected, where $y$ is the fractional energy transferred to the 
hadronic final state and $\theta_{ele}$ is
the polar angle of the scattered electron in the $ep$ reaction. The $y=1$
line indicates the kinematic limit for HERA running with 920 GeV proton.}
\label{fig:4}
\end{figure}

Figure \ref{fig:4} shows the distribution in $x$ and $Q^2$ 
of selected events containing at least one 
pair of $K_s^0$ candidates. The virtual photon-proton 
center of mass energy is in the range \mbox{$50<W<250$ GeV}.

\begin{figure}[htb]
%\framebox[55mm]{\includegraphics{wxsecbw.eps}}{\rule[-21mm]{0mm}{43mm}}
\epsfig{file=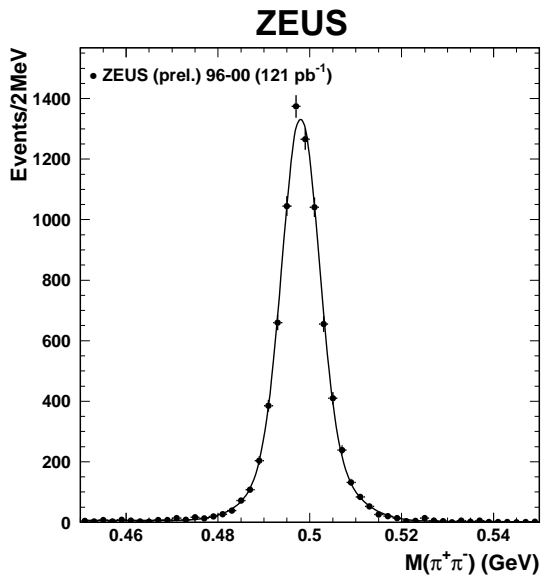,width=0.5\textwidth}
\caption{The distribution of $\pi^+\pi^-$ invariant-mass 
events with two $K_s^0$ candidates 
passing all selection cuts, The solid line shows the result of a
fit using one linear and two Gaussian functions.}
\label{fig:5}
\end{figure}

Figure~\ref{fig:5} shows the $M(\pi^+\pi^-)$ distribution in the range
\mbox{$0.45<M(\pi^+\pi^-)<0.55$ GeV} after the $K_s^0$ pair candidate selection.

%When analyzing the $K_s^0K_s^0$ spectrum, one has to be aware of a 
%possible strong enhancement near the $K_s^0K_s^0$ threshold. Since
%the high $K_s^0K_s^0$ mass is the region of interest for this analysis,
%the complication due to the threshold region is avoided by requiring
A cut \mbox{$cos\theta_{K_s^0K_s^0}<0.92$} is applied to the
selected $K_s^0$ pair candidates, where $\theta_{K_s^0K_s^0}$ 
is the opening angle between the two $K_s^0$ candidates in the 
laboratory frame. This cut removes phase-space effects and 
the presence of $f_0(980)$/$a_0(980)$ at the threshold, which decays to 
collinear $K_s^0$ pairs affecting the measurement in the 1300 MeV mass region.

\begin{figure}[htb]
%\framebox[55mm]{\includegraphics{wxsecbw.eps}}{\rule[-21mm]{0mm}{43mm}}
\epsfig{file=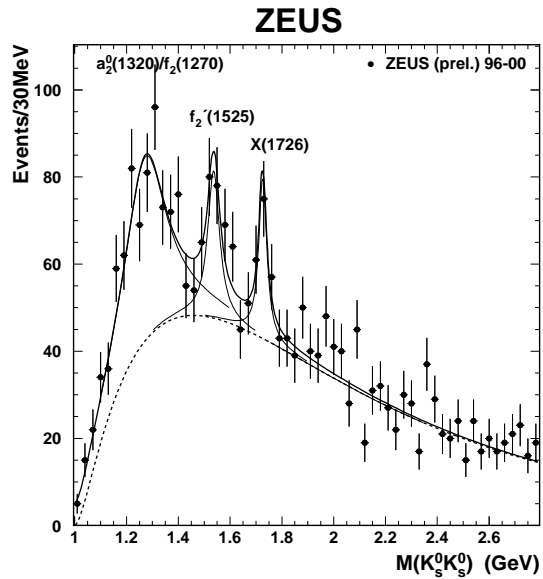,width=0.5\textwidth}
\caption{The $K_s^0K_s^0$ invariant-mass spectrum. The solid line is the result 
of a fit using three Breit-Wigner and a background function.}
\label{fig:6}
\end{figure}

Figure \ref{fig:6} shows the $K_s^0K_s^0$ 
invariant-mass spectrum in the range 
\mbox{$0.995~<M_{K_s^0K_s^0}~<2.795$} GeV 
for data with \mbox{$cos\theta_{K_s^0K_s^0}<0.92$} 
(filled circles with error bar). Two clear 
peaks are seen, one around 1500 MeV, consistent with the
well-established $f_2^\prime(1525)$~\cite{pdg}, and the other 
around 1700 MeV, close to $f_0(1710)$~\cite{pdg}. There is 
also an enhancement around 1300 MeV, consistent with 
the $f_2(1270)/a_2(1320)$ interference.

It was found that most of the $K_s^0$ pair candidates selected after all
cuts are in the ``gluon-rich'' region $x_{p}=2p_B/Q > 1$ of
the target region of the Breit frame, which 
corresponds to the remnant of the proton \cite{breit1,breit2} , 
where $p_B$ is the absolute momentum of the $K_s^0K_s^0$ in the Breit frame.

\subsection{Summary}

The first observation in deep inelastic $ep$ scattering of a state 
near $1525$ MeV, consistent with 
$f^\prime_2(1525)$, 
and another close to $f_0(1710)$ is reported. 

An enhancement which can be interpreted as the $f_2(1270)$/$a_2^0(1320)$ 
interference 
is observed, but its measurement is affected by 
the presence of the $f_0(980)$/$a_0(980)$ at the $K_s^0K_s^0$ threshold.

%, fitted with two Guassians and
%a linear function. 
%The linear function fits the background, one of the
%Gaussians fits the peak region in the central $\pi^+\pi^-$ invariant
%mass distribution and the other Gaussian improves the fit at the tails.
%Only $K_s^0$ candidates 
%and the invariant mass of the $K_s^0$ pair candidate $M(K_s^0,K_s^0)$
%was reconstructed 

\end{document}